\documentclass{IEEEtran}
\usepackage{amsmath}
\usepackage{amsfonts}
\usepackage{amsthm}
\usepackage{amssymb}
\usepackage{cite}
\usepackage{url}
\usepackage{subfigure}
\usepackage{graphicx}
\usepackage{color}
\newcommand{\ie}{{\em i.e., }}
\newcommand{\eg}{{\em e.g., }}

\newcommand{\bl}{\left(}
\newcommand{\br}{\right)}
\newcommand{\blc}{\left\{}
\newcommand{\brc}{\right\}}

\renewcommand{\vec}[1]{\mathbf{#1}}

\newcommand{\sign}{\mathop{\rm sgn}}

\newcommand{\beq} {\begin{equation}}
\newcommand{\eeq}[1] {\label{#1}\end{equation}}
\newcommand{\bfg} {\begin{figure}\begin{center}}
\newcommand{\efg}[1] {\end{center}\label{#1}\end{figure}}

\renewcommand{\vec}[1]{\mathbf{#1}}

\newcommand{\bw}{{\bf w}}
\newcommand{\hw}{{\hat{\bf w}}}

\newcommand{\bx}{{\bf x}}

\newcommand{\bbi}{{\bf I}}

\newtheorem{thm}{Theorem}

\begin{document}

\title{Regularized Least-Mean-Square Algorithms}

\author{Yilun~Chen,~\IEEEmembership{Student Member,~IEEE,}
Yuantao~Gu,~\IEEEmembership{Member,~IEEE,}
and~Alfred~O.~Hero,~III,~\IEEEmembership{Fellow,~IEEE}
\thanks{Y. Chen and A. O. Hero are with the Department of Electrical Engineering
and Computer Science, University of Michigan, Ann Arbor, MI 48109,
USA. Tel: 1-734-763-0564. Fax: 1-734-763-8041. Emails: \{yilun, hero\}@umich.edu.}

\thanks{Y. Gu is with the Department of Electronic Engineering, Tsinghua University, Beijing 100084, China. Tel:+86-10-62792782, Fax: +86-10-62770317. Email: gyt@tsinghua.edu.cn.}

\thanks{This work was partially supported by AFOSR, grant number FA9550-06-
1-0324.
}}

\maketitle
\begin{abstract}
We consider adaptive system identification problems with convex constraints and propose a family of regularized Least-Mean-Square (LMS) algorithms. We show that with a properly selected regularization parameter the regularized LMS provably dominates its conventional counterpart in terms of mean square deviations. We establish simple and closed-form expressions for choosing this regularization parameter.  For identifying an unknown sparse system we propose sparse and group-sparse LMS algorithms, which are special examples of the regularized LMS family.  Simulation results demonstrate the advantages
of the proposed filters in both convergence rate and steady-state
error under sparsity assumptions on the true coefficient vector.
\end{abstract}
\section{Introduction}
The Least Mean Square (LMS) algorithm, introduced by Widrow and
Hoff \cite{Widrow85}, is a popular method for adaptive system identification.
Its applications include echo cancelation, channel equalization,
interference cancelation and so forth. Although there exist
algorithms with faster convergence rates such as the Recursive
Least Square (RLS) methods, LMS-type methods are popular because
of its ease of implementation, low computational costs and
robustness.

In many scenarios often prior information about
the unknown system is available. One important example is when the impulse response of the unknown
system is known to be sparse, containing only a few large
coefficients interspersed among many small ones. Exploiting
such prior information can improve the filtering performance and
has been investigated for several years. Early work includes heuristic
online selection of active taps \cite{Kawamura86, Homer98, li2006parallel} and sequential partial updating
\cite{Etter85, Hero05}; other algorithms assign proportional step sizes of different
taps according to their magnitudes, such as the Proportionate
Normalized LMS (PNLMS) and its variations \cite{Gay98, Duttweiler00}.

Motivated by LASSO \cite{Tibshirani96} and recent progress in compressive sensing
\cite{Candes06, Baraniuk07}, the authors in \cite{chen2009sparse} introduced an $\ell_1$-type regularization
to the LMS framework resulting in two sparse LMS methods called ZA-LMS and RZA-LMS. This methodology was also
applied to other adaptive filtering frameworks such as RLS \cite{babadi2010sparls, angelosante2010online} and
projection-based adaptive algorithms \cite{kopsinis2010online}. Inheriting the
advantages of conventional LMS methods such as robustness and low computational costs, the sparse LMS  filters were empirically demonstrated to achieve superior performances in both convergence rate and steady-state behavior, compared to the standard LMS when the system is sparse. However, while the regularization parameter needs to be tuned there is no systematical way to choose the parameter. Furthermore, the analysis of \cite{chen2009sparse} is only based on the $ \ell_1 $ penalty and not applicable to other regularization schemes.

In this paper,  we extend the methods presented in \cite{chen2009sparse, gu2009l_}  to a  broad family  of regularization penalties and consider LMS and Normalized LMS algorithms (NLMS) \cite{Widrow85} under general convex constraints.  In addition, we allow the convex constraints to be time-varying. This results in a regularized LMS/NLMS\footnote{We treat NLMS as a special case of the general LMS algorithm and will not distinguish the two unless required for clarity.} update equation with an additional sub-gradient term. We show that the regularized LMS provably dominates its conventional counterpart if a proper regularization parameter is selected. We also establish a simple and closed-form formula to choose this parameter. For white input signals, the proposed parameter selection guarantees dominance of the regularized LMS over the conventional LMS. Next, we show that the sparse LMS filters in \cite{chen2009sparse},  \ie ZA-LMS and RZA-LMS, can be obtained as special cases of the regularized LMS family introduced here. Furthermore, we consider a group-sparse adaptive FIR filter response that is useful for practical applications \cite{Duttweiler00, schreiber2002advanced}. To enforce group sparsity  we use $\ell_{1,2}$ type regularization functions \cite{yuan2006model} in the regularized LMS framework.  For sparse and group-sparse LMS methods, we propose alternative closed-form expressions for selecting the regularization parameters. This guarantees provable dominance for both white and correlated input signals. Finally, we demonstrate performance advantages of our proposed sparse and group-sparse LMS filters using numerical simulation. In particular, we show that the regularized LMS method is robust to model mis-specification and outperforms the contemporary projection based methods \cite{kopsinis2010online} for equivalent computational cost.

The paper is organized as follows. Section II formulates the problem and introduces the regularized LMS algorithm. In Section III we develop LMS filters for sparse and group-sparse system identification. Section IV provides numerical simulation results and Section V summarizes our principal conclusion. The proofs of theorems are provided in the Appendix.

\emph{Notations}: In the following parts of paper, matrices and vectors
are denoted by boldface upper case letters and boldface lower case
letters, respectively; $(\cdot)^T$ denotes the transpose
operator, and $ \|\cdot\|_1 $ and $ \|\cdot\|_2 $ denote the $ \ell_1 $ and $ \ell_2 $ norm of a vector, respectively.

\section{Regularized LMS}
\subsection{LMS framework}
We begin by briefly reviewing the framework of the LMS filter, which forms the basis of our derivations to follow.
Denote the coefficient vector and the input signal vector of the
adaptive filter as
\begin{equation}
	\label{eq:hwn}
\hw_n =
[\hat w_{n,0}, \hat w_{n,1}, \cdots, \hat w_{n,N-1}]^T
\end{equation}
and
\begin{equation}
\bx_n = [x_n,x_{n-1},\cdots,x_{n-N+1}]^T,
\end{equation}
respectively, where $n$ is the
time index, $x_n$ is the input signal, $\hat w_{n,i}$ is the $i$-th coefficient at time $n$ and $N$ is the length of
the filter. The goal of the LMS algorithm is to identify the true
system impulse response $\bw$ from the input signal $x_n$ and
the desired output signal $y_n$, where
\begin{equation}
  y_n = \bw^T\vec{x}_n+v_n.
\end{equation}
$v_n$ is the observation noise which is assumed to be
independent with $x_n$.

Let $e_n$ denote the instantaneous error between the filter
output $\hw^T_n\vec{x}_n$ and the desired output $y_n$:
\begin{equation}
  e_n = y_n-\hw^T_n\vec{x}_n.
\end{equation}
In the standard LMS framework, the cost function $L_n$ is defined
as the instantaneous square error
\begin{displaymath}
  L_n(\hw_n) = \frac{1}{2}e^2_n
\end{displaymath}
and the filter coefficient vector is updated in a stochastic gradient descent manner:
\begin{equation}
	\label{eq:lms}
  \hw_{n+1} = \hw_n- \mu_{n} \nabla L_n(\bw_n) =  \hw_n+
  \mu_{n} e_n\vec{x}_n,
\end{equation}
where $\mu_{n}$ is the step size controlling the convergence and the
steady-state behavior of the LMS algorithm. We refer to (\ref{eq:lms}) as the conventional LMS algorithm and emphasize that $ \mu_{n} $ can be both time-varying and functions of $ \bx_n $. For example,
\begin{equation}
	\label{eq:nlms_stepsize}
\mu_{n} = \frac{\alpha_n}{\|\bx_n\|_2^2}
\end{equation}
yields the normalized LMS (NLMS) algorithm with variable step size $ \alpha_n $.

\subsection{Regularized LMS}
Conventional LMS algorithms do not impose any model on the true system response $\bw$. However, in practical scenarios often prior knowledge of $ \bw $ is available. For example, if the system is known to be sparse, the $ \ell_1 $ norm of $ \bw $ can be upper bounded by some constant \cite{Tibshirani96}. In this work, we study the adaptive system identification problem where the true system is constrained by
\begin{equation}
	\label{eq:def_fn}
f_n(\bw) \le \eta_n,
\end{equation}
where $ f_n(\cdot) $ is a convex function and $ \eta_n $ is a constant. We note that the subscript $ n $ in $ f_n(\cdot) $ allows adaptive constraints that can vary in time. Based on (\ref{eq:def_fn}) we propose a regularized instantaneous cost function
\begin{equation}
L_{n}^{\text{reg}}(\hw_n) = \frac{1}{2}e_{n}^{2} + \gamma_n f_{n}(\hw_n)
\end{equation}
and update the coefficient vector by
\begin{equation}
	\label{eq:rlms}
\begin{aligned}
\hw_{n+1} & = \hw_{n} - \mu_{n} \nabla L_{n}^{\text{reg}}(\hw_n)\\
					  &= \hw_{n} + \mu_n e_n \bx_n - \rho_n \partial f_n(\hw_n),
\end{aligned}
\end{equation}
where $ \partial f_n(\cdot) $ is the sub-gradient of the convex function $ f_n(\cdot) $, $ \gamma_n $ is the regularization parameter and $ \rho_n  = \gamma_n \mu_n$.

Eq. (\ref{eq:rlms}) is the proposed regularized LMS. Compared to its conventional counterpart, the regularization term, $ -\rho_n\partial f_n(\hw_n) $,  always promotes the coefficient vector to satisfy the constraint (\ref{eq:def_fn}). The parameter $ \rho_n $ is referred to as the regularization step size. Instead of tuning $ \rho_n $ in an \textit{ad hoc} manner, we establish a systematic approach to choosing $ \rho_n $.

\begin{thm}
\label{thm:1}
Assume both $ \{x_n\} $ and $ \{v_n\} $ are Gaussian independent and identically distributed (i.i.d.) processes that are mutually independent.  For any $ n>1 $
\begin{equation}
    \label{eq:strongdominance}
E\left\| {{\bf{\hat w}}_n - {\bf{w}}} \right\|_2^2 \le E\left\| {{\bf{\hat w}}_n' - {\bf{w}}} \right\|_2^2
\end{equation}
if $ \hw_0 = \hw_0'$ and $ \rho_n \in [0, 2\rho_n^*] $, where $ \bw $ is the true coefficient vector and $ \hw_n' $ and  $ \hw_n $ are filter coefficients updated by (\ref{eq:lms}) and (\ref{eq:rlms}) with the same step size $ \mu_n $, respectively. $ \rho_n^* $ is calculated by
\begin{equation}
	\label{eq:rholms}
\rho_n^* = \max\blc(1-\mu_n\sigma_x^2)\frac{f_n(\hw_n)-\eta_n}{\|\partial
f_n(\hw_n)\|_2^2},0\brc
\end{equation}
if $\mu_n $ are constant values (LMS), or
\begin{equation}
	\label{eq:rhonlms}
\rho_n^* = \max\blc(1-\alpha_n/N)\frac{f_n(\hw_n)-\eta_n}{\|\partial
f_n(\hw_n)\|_2^2},0\brc
\end{equation}
if $ \mu_n $ is chosen using  (\ref{eq:nlms_stepsize}) (NLMS), where $ N $ is the filter length, $\sigma_x^2$ is the variance of $\{x_n\}$ and $ \eta_n $ is an upper bound of $ f_n(\bw) $ defined  in (\ref{eq:def_fn}).
\end{thm}
The proof of Theorem \ref{thm:1} is provided in the Appendix.

\textit{Remark 1}.
Theorem \ref{thm:1} shows that with the same initial condition and step size $ \mu_n $, the regularized LMS algorithm provably dominates conventional LMS when the input signal is white. The parameter $ \rho_n^* $ in (\ref{eq:rholms}) or (\ref{eq:rhonlms}) can be used as the value for $ \rho_n $  in (\ref{eq:rlms}) to guarantee that regularized LMS will have lower MSD than conventional LMS. The value $\rho_n^*$ only requires specification of the noise variance and $ \eta_n $ which upper bounds the true value $ f_n(\bw) $. Simulations in latter sections show that the performance of the regularized LMS is robust to misspecified values of $ \eta_n $.

\textit{Remark 2}.
Eq. (\ref{eq:rholms}) and (\ref{eq:rhonlms}) indicate that to ensure superiority the regularization is only ``triggered" if $ f_n(\hw_n) > \eta_n $. When $ f_n(\hw_n) \le \eta_n $, $ \rho_n^* = 0 $ and the regularized LMS reduces to the conventional LMS.

\textit{Remark 3}.
The closed form expression for $ \rho_n^* $ is derived based on the white input assumption. Simulation results in latter sections show that the (\ref{eq:rholms}) and (\ref{eq:rhonlms}) are also empirically good choices even for correlated input signals. Indeed, in the next section we will show that provable dominance can be guaranteed for correlated inputs when the regularization function is suitably selected.

\section{Sparse system identification}

\begin{figure}
\begin{center}
\includegraphics[scale=0.5]{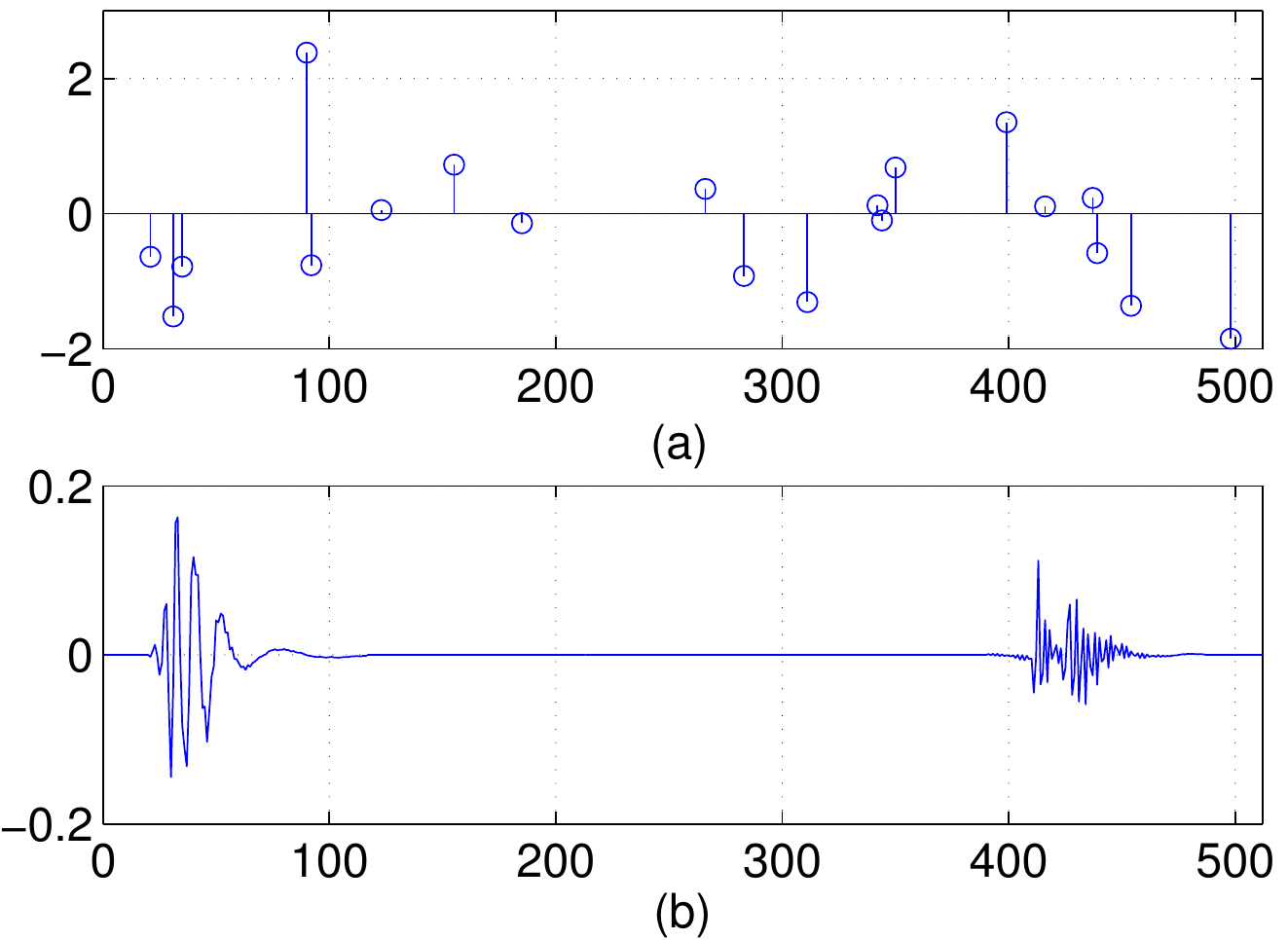}
\caption{Examples of (a) a general sparse system and  (b) a group-sparse system.}
\label{fig:1}
\end{center}
\end{figure}

A sparse system contains only a few large coefficients interspersed among many negligible ones. Such sparse systems are arise in many applications such as digital TV transmission channels \cite{schreiber2002advanced} and acoustic echo channels \cite{Duttweiler00}. Sparse systems can be further divided into general sparse systems and group-sparse systems, as shown in Fig. \ref{fig:1} (a) and Fig. \ref{fig:1} (b), respectively. Here we apply our regularized LMS to both general and group sparse system identification. We show that ZA-LMS and RZA-LMS in \cite{chen2009sparse} are special examples of regularized LMS. We then propose group-sparse LMS algorithms for identifying group-sparse systems.

\subsection{Sparse LMS}
For a general sparse system, the locations of active non-zero coefficients are unknown but one may know an upper bound on their number. Specifically, we will assume that the impulse response $ \bw $ satisfies
\begin{equation}
\|\bw\|_0 \le k,
\end{equation}
where $ \|\cdot\|_0 $ is the $ \ell_0 $ norm denoting the number of non-zero entries of a vector, and $ k $ is a known upper bound. As  the $ \ell_0 $ norm is non-convex it is not suited to the proposed framework. Following \cite{Tibshirani96} and \cite{Candes06}, we instead adopt the $ \ell_1 $ norm as a surrogate approximation to the $ \ell_0 $ norm:
\begin{equation}
\|\bw\|_1 = \sum_{i=0}^{N-1} |w_i| .
\end{equation}
Using the  regularization penalty $f_n(\mathbf w)=\|\mathbf w\|_1$ in regularized LMS (\ref{eq:rlms}), we obtain
\begin{equation}
	\label{eq:zalms}
\hw_{n+1} = \hw_{n} + \mu_n e_n \bx_n - \rho_n \sign{\hw_n},
\end{equation}
where the  component-wise $ \sign(\cdot) $ function is defined as
\begin{equation}
  \sign(x) =
  \begin{cases}
    x/|x| & x \ne 0\\
    0 & x=0
  \end{cases}.
\end{equation}
Equation (\ref{eq:zalms}) yields the ZA-LMS introduced in \cite{chen2009sparse}. The regularization parameter $ \rho_n $ can be calculated by (\ref{eq:rholms}) for LMS and by (\ref{eq:rhonlms}) for NLMS, where $ f_n(\hw_n) = \|\hw_n\|_1 $ and $ \eta_n $ is an estimate of the true $ \|\bw\|_1 $.

An alternative approach to approximating the $ \ell_0 $ norm is to consider the following function \cite{candes2008enhancing, chen2009sparse, kopsinis2010online}:
\begin{equation}
	\label{eq:appl0}
\|\bw\|_0 \simeq \sum_{i=0}^{N-1} \frac{1}{|w_i|+\delta} \cdot |w_i|,
\end{equation}
where $ \delta $ is a sufficiently small positive real number. Interpreting (\ref{eq:appl0}) as a weighted $ \ell_1 $ approximation, we propose the regularization function $ f_n(\bw)  $
\begin{equation}
	\label{eq:wl1}
f_n(\bw) = \sum_{i=0}^{N-1} \beta_{n,i} \cdot |w_i|,
\end{equation}
and
\begin{equation}
    \beta_{n,i} = \frac{1}{|\hat w_{n,i}|+\delta},
\end{equation}
where $ \hat w_{n,i} $ is the $ i $-th coefficient of $ \hw_n $ defined in (\ref{eq:hwn}). Using (\ref{eq:wl1}) in (\ref{eq:rlms}) yields
\begin{equation}
    \label{eq:rzalms}
\hat w_{n+1,i} = \hat w_{n,i} + \mu_n e_n x_{n-i} - \rho_n \beta_{n,i} ~{\sign{\hat w_{n,i}}},
\end{equation}
which is a component-wise update of the RZA-LMS proposed in \cite{chen2009sparse}. Again, $\rho_n$ can be computed using (\ref{eq:rholms}) for LMS or (\ref{eq:rhonlms}) for NLMS, where $\eta_n$ is an estimate of the true $ \|\bw\|_0 $, \ie the number of the non-zero coefficients.

\subsection{Group-sparse LMS}
\label{sec:groupsparselms}
In many practical applications, a sparse system often exhibits a grouping structure, \ie coefficients in the same group are highly correlated and take on the values zero or non-zero as a group, as shown in Fig. \ref{fig:1} (b). The motivation for developing group-sparse LMS is to take advantage of such a structure.

We begin by employing the mixed $ \ell_{1,2} $ norm for promoting group-sparsity, which was originally proposed in \cite{yuan2006model} and has been widely adopted for various structured sparse regression problems \cite{meier2008group, bach2008consistency}. The $ \ell_{1,2} $ norm of a vector $ \bw $  is defined as
\begin{equation}
\|\bw\|_{1,2} = \sum_{j=1}^{J} \|\bw_{I_j}\|_2,
\end{equation}
where $ \{I_j\}_{j=1}^J $ is a group partition of the whole index set $ I = \{0,1,\ldots,N-1\} $:
\begin{equation}
\bigcup_{j=1}^J I_j = I, \quad I_j  \cap I_{j'} = \phi \text{  when } j \ne j',
\end{equation}
and $ \bw_{I_j} $ is a sub-vector of $ \bw $ indexed by $ I_j $. The $ \ell_{1,2} $ norm is a mixed norm: it encourages correlation among coefficients inside each group
via the $\ell_2$ norm and promotes sparsity across those groups using the $\ell_1$ norm. $\|\bw\|_{1,2}$ is convex in $\bw$ and reduces to $\|\bw\|_1$ when each group contains only one coefficient, \ie
\begin{equation}
    |I_1| = |I_2| = \cdots = |I_J| = 1,
\end{equation}
where $|\cdot|$ denotes the cardinality of a set. Employing $f_n(\bw) = \|\bw\|_{1,2}$, the $\ell_{1,2}$ regularized LMS, which we refer to as GZA-LMS, is
\begin{equation}\label{eq:l12rlms}
    \hw_{n+1,I_j} = \hw_{n,I_j} + \mu_n e_n \bx_{I_j} - \rho_n \frac{\hw_{n,I_j}}{\|\hw_{n,I_j\|_2}+\delta}, ~~ j=1,...,J,
\end{equation}
and $ \delta $ is a sufficiently small number ensuring a non-zero denominator.
To the best of our knowledge this is the first time that the $ \ell_{1,2} $ norm has been proposed for the LMS adaptive filters.

\begin{figure}
\begin{center}
\includegraphics[scale=0.92]{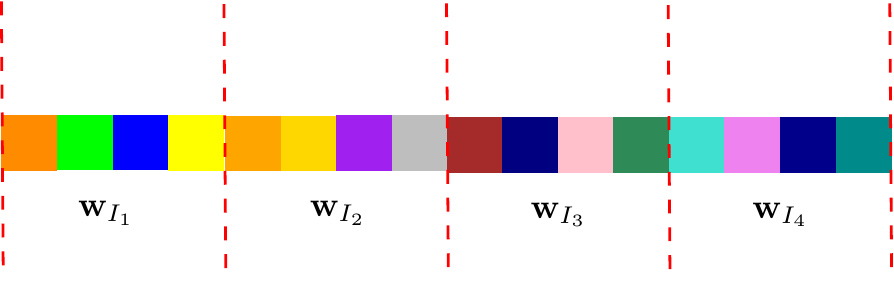}
\caption{A toy example illustrating the $ \ell_{1,2} $ norm of a $ 16 \times 1 $ coefficient vector $ \bw $: $ \|\bw\|_{1,2} = \sum_{j=1}^4 \|\bw_{I_j}\|_2 $.}
\label{fig:2}
\end{center}
\end{figure}

To further promote group selection we consider the following weighted $\ell_{1,2}$ regularization as a group-wise generalization of (\ref{eq:wl1}):
\begin{equation}
    \label{eq:wl12}
  f_n(\bw) = \sum_{j=1}^J \beta_{n,j} \|\bw_{I_j}\|_2,
\end{equation}
where $ \beta_{n,j} $ is a re-weighting parameter defined by
\begin{equation}
    \beta_{n,j} = \frac{1}{\|\hw_{n,I_j}\|_2 + \delta},
\end{equation}
and the corresponding regularized LMS update is then
\begin{equation}
  \label{eq:wl12lms}
   \hw_{n+1,I_j} = \hw_{n,I_j} + \mu_n e_n \bx_{I_j} - \rho_n  \beta_{n,j} \frac{\hw_{n,I_j}}{\|\hw_{n,I_j\|_2}+\delta}, ~~ j=1,...,J,
\end{equation}
which is referred to as GRZA-LMS.

As both the $\ell_{1,2}$ norm and the weighted $\ell_{1,2}$ norm are convex, Theorem \ref{thm:1} applies under the assumption of white input signals and $\rho_n$ can be calculated by (\ref{eq:rholms}) or (\ref{eq:rhonlms}). The parameter $\eta_n$ can be chosen as an estimate of the true $\|\bw\|_{1,2}$ for GZA-LMS (\ref{eq:l12rlms}), or the number of non-zero groups of $\bw$ for GRZA-LMS (\ref{eq:wl12lms}).


Finally, we note that GZA-LMS and GRZA-LMS reduce to ZA-LMS and RZA-LMS, respectively, if each group contains only one element.

\subsection{Choosing regularization parameter for correlated input}
Theorem \ref{thm:1} gives a closed form expression for $\rho_n$ and (\ref{eq:rholms}) or (\ref{eq:rhonlms}) is applicable for any convex $f_n(\bw)$. However, the dominance over conventional LMS is only guaranteed when the input signal is white. Here we develop an alternative formula to determine $\rho_n$ that applies to correlated input signals for sparse and group-sparse LMS, \ie (\ref{eq:zalms}), (\ref{eq:rzalms}), (\ref{eq:l12rlms}) and (\ref{eq:wl12lms}).

We begin by considering the weighted $\ell_{1,2}$ regularization (\ref{eq:wl12}) and the corresponding GRZA-LMS update (\ref{eq:wl12lms}). Indeed, the other three algorithms, \ie (\ref{eq:l12rlms}), (\ref{eq:rzalms}) and (\ref{eq:zalms}), can be treated as special cases of (\ref{eq:wl12lms}). For general wide-sense stationary (WSS) input signals, the regularization parameter $\rho_n$ of (\ref{eq:wl12lms}) can be selected according the following theorem.
\begin{thm}
\label{thm:2}
Assume $\{x_n\}$ and $\{v_n\}$ are WSS stochastic processes which are mutually independent. Let $\hw_{n}$ and $\hw_n'$ be filter coefficients updated by (\ref{eq:wl12lms}) and (\ref{eq:lms}) with the same $\mu_n$, respectively. Then,
\begin{equation}\label{eq:weakdomination}
    E\left\| {{\bf{\hat w}}_{n+1} - {\bf{w}}} \right\|_2^2 \le E\left\| {{\bf{\hat w}}_{n+1}' - {\bf{w}}} \right\|_2^2
\end{equation}
if $ \hw_n = \hw_n'$ and $ \rho_n \in [0, 2\rho_n^*] $, $ \bw $ is the true coefficient vector and $ \rho_n^* $ is
\begin{equation}
    \label{eq:wl12rho}
\rho_n^* = \max\blc\frac{f_n(\hw_n)-\eta_n - \mu_n r_n}{\|\partial
f_n(\hw_n)\|_2^2},0\brc,
\end{equation}
where $f_n(\hw_n)$ is determined by (\ref{eq:wl12}), $ \eta_n $ is an upper bound of $ f_n(\bw)$ and
\begin{equation}
    \label{eq:def_rn}
    r_n = \hw_n^T \bx_n \cdot \bx_n^T \partial f_n(\hw_n) + \eta_n \cdot \max_j\blc\frac{\|\bx_{I_j}\|_2}{\beta_{n,j}}\brc\cdot |\bx_n^T\partial f_n(\hw_n)|.
\end{equation}
\end{thm}
The proof of Theorem \ref{thm:2} can be found in the Appendix. We make the following remarks.

\emph{Remark 4}. Theorem \ref{thm:2} is derived from the general form (\ref{eq:wl12lms}) and can be directly specialized to (\ref{eq:l12rlms}), (\ref{eq:rzalms}) and (\ref{eq:zalms}). Specifically,
\begin{itemize}
\item GZA-LMS (\ref{eq:l12rlms}) can be obtained by assigning $\beta_{n,j} = 1$;
\item RZA-LMS (\ref{eq:rzalms}) can be obtained when $|I_j| = 1, j=1,...,J$;
\item ZA-LMS (\ref{eq:zalms}) can be obtained when both $|I_j| = 1, j=1,...,J$ and $\beta_{n,j} = 1$.
\end{itemize}

\emph{Remark 5.} Theorem \ref{thm:2} is valid for any WSS input signals. However, the dominance result in (\ref{eq:weakdomination}) is weaker than that in Theorem \ref{thm:1}, as it requires  $ \hw_n = \hw_n'$ at each iteration.

\emph{Remark 6.} Eq. (\ref{eq:wl12rho}) can be applied to both LMS and NLMS, depending on if $\mu_n$ are deterministic functions of $\bx_n$ as specified in (\ref{eq:nlms_stepsize}). This is different from Theorem \ref{thm:1} where we have separate expressions for LMS and NLMS.

\emph{Remark 7.} $\rho_n^*$ in (\ref{eq:wl12rho}) is non-zero only if $f_n(\hw_n)$ is greater than $\eta_n + \mu_n r_n$ (rather than $\eta_n$ as presented in Theorem \ref{thm:1}). This may yield a more conservative performance.
\section{Numerical simulations}
In this section we demonstrate our proposed sparse LMS algorithms by  numerical simulations. Multiple experiments are designed to evaluate their performances over a wide range of conditions.

\subsection{Identifying a general sparse system}
\begin{figure}
\begin{center}
  \includegraphics[width=7cm]{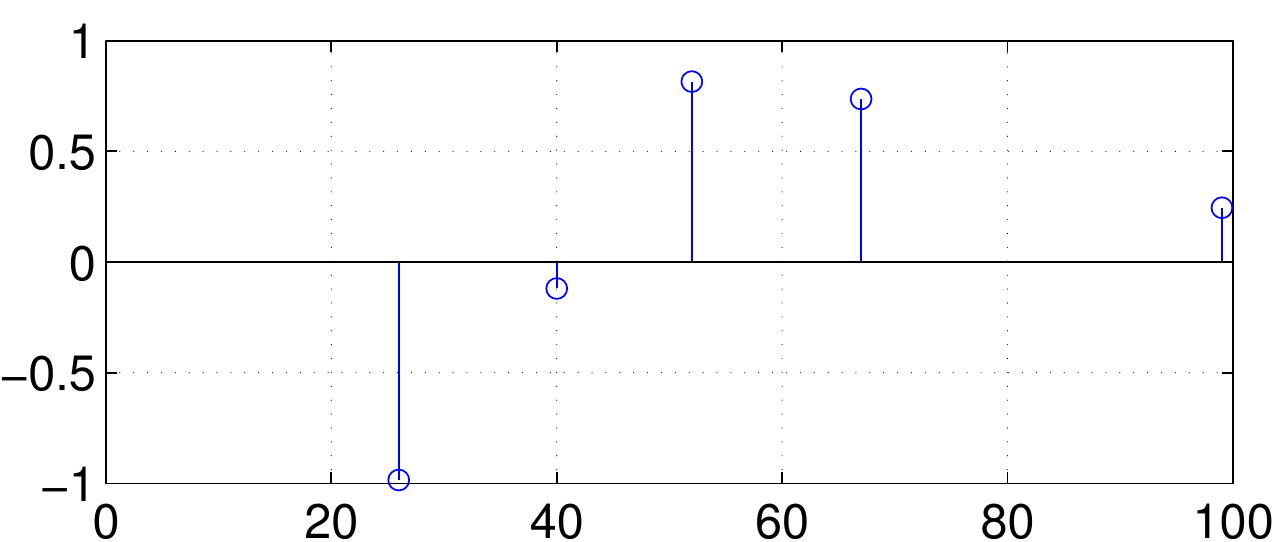}\\
  \caption{The general sparse system used for simulations.}\label{fig:3}
\end{center}
\end{figure}

Here we perform evaluation of the proposed filters for general sparse system identification, as illustrated in Fig. \ref{fig:1} (a). There are 100 coefficients in the time varying system and only five of them are non-zero. The five non-zero coefficients are assigned to random locations and their values are also randomly drawn from a standard Gaussian distribution. The resultant true coefficient vector is plotted in Fig. \ref{fig:3}.

\subsubsection{White input signals}
\label{sec:whiteinput}
\begin{figure}
\begin{center}
  \includegraphics[width=7cm]{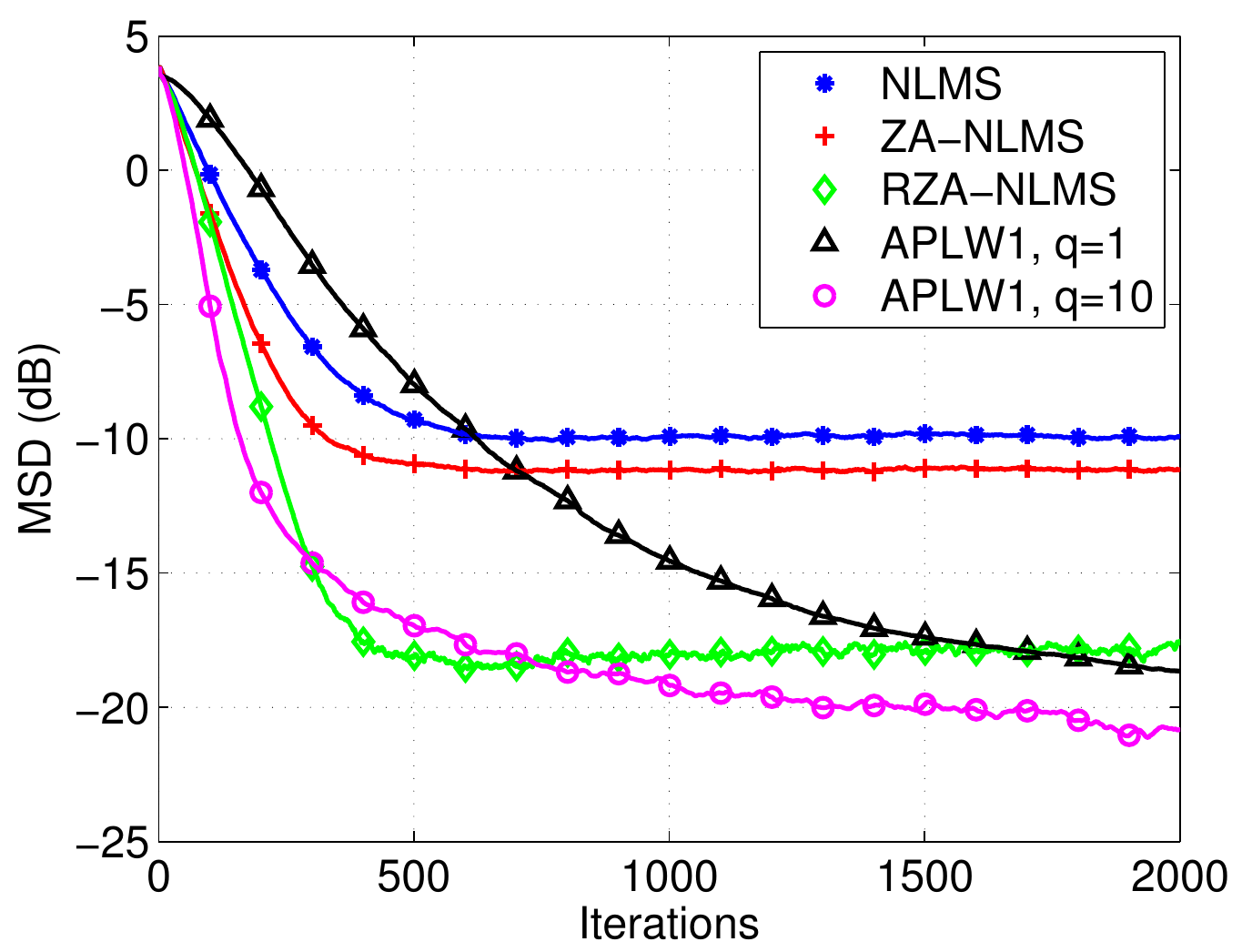}\\
  \caption{White input signals: performance comparison for different filters.}\label{fig:4}
\end{center}
\end{figure}

Initially we simulate white Gaussian input signal $\{x_n\}$ with zero mean and unit variance. The measurement noise $\{v_n\}$ is an independent Gaussian random process of zero mean and variance $\sigma_v^2 = 0.1$. For ease of parameter selection, we implement NLMS-type filters in our simulation. Three filters (NLMS, ZA-NLMS and RZA-NLMS) are implemented and their common step-size $\mu_n$ is set via (\ref{eq:nlms_stepsize}) with $\alpha_n = 1$. The regularization parameter $\rho_n$ is computed using (\ref{eq:rhonlms}), where $\eta_n$ is set to $\eta_n = \|\bw\|_1$ (\ie the true value) for ZA-NLMS and $\eta_n = 5$ for RZA-NLMS. For comparison we also implement a recently proposed sparse adaptive filter, referred to as APWL1 \cite{kopsinis2010online}, which sequentially projects the coefficient vector onto weighted $\ell_1$ balls. We note that our simulation setting is identical to that used in \cite{kopsinis2010online} and thus we adopt the same tuning parameters for APWL1. In addition, the weights $\beta_{n,i}$ for RZA-NLMS is scheduled in the same manner as that in \cite{kopsinis2010online} for a fair comparison. The simulations are run 100 times and the average estimates of mean square deviation (MSD) are shown in Fig. \ref{fig:4}.

It can be observed that ZA-NLMS improves upon NLMS in both convergence rate and steady-state behavior and RZA-NLMS does even better. The parameter $q$ of APLW1 is the number of samples used in each iteration. One can see that RZA-NLMS outperforms APLW1 when $q=1$, \ie the case that APLW1 operates with the same memory storage as RZA-NLMS. With larger $p$ APLW1 begins to perform better and exceeds RZA-NLMS when $q \ge 10$. However, there is a trade-off between the system complexity and filtering performance, as APWL1 requires $\mathcal{O}(qN)$ for memory storage and $\mathcal{O}(N \log_2 N + qN)$ for computation, in contrast to LMS-type methods which require only $\mathcal{O}(N)$ for both memory and computation.

\begin{figure}
\begin{center}
  \includegraphics[width=7cm]{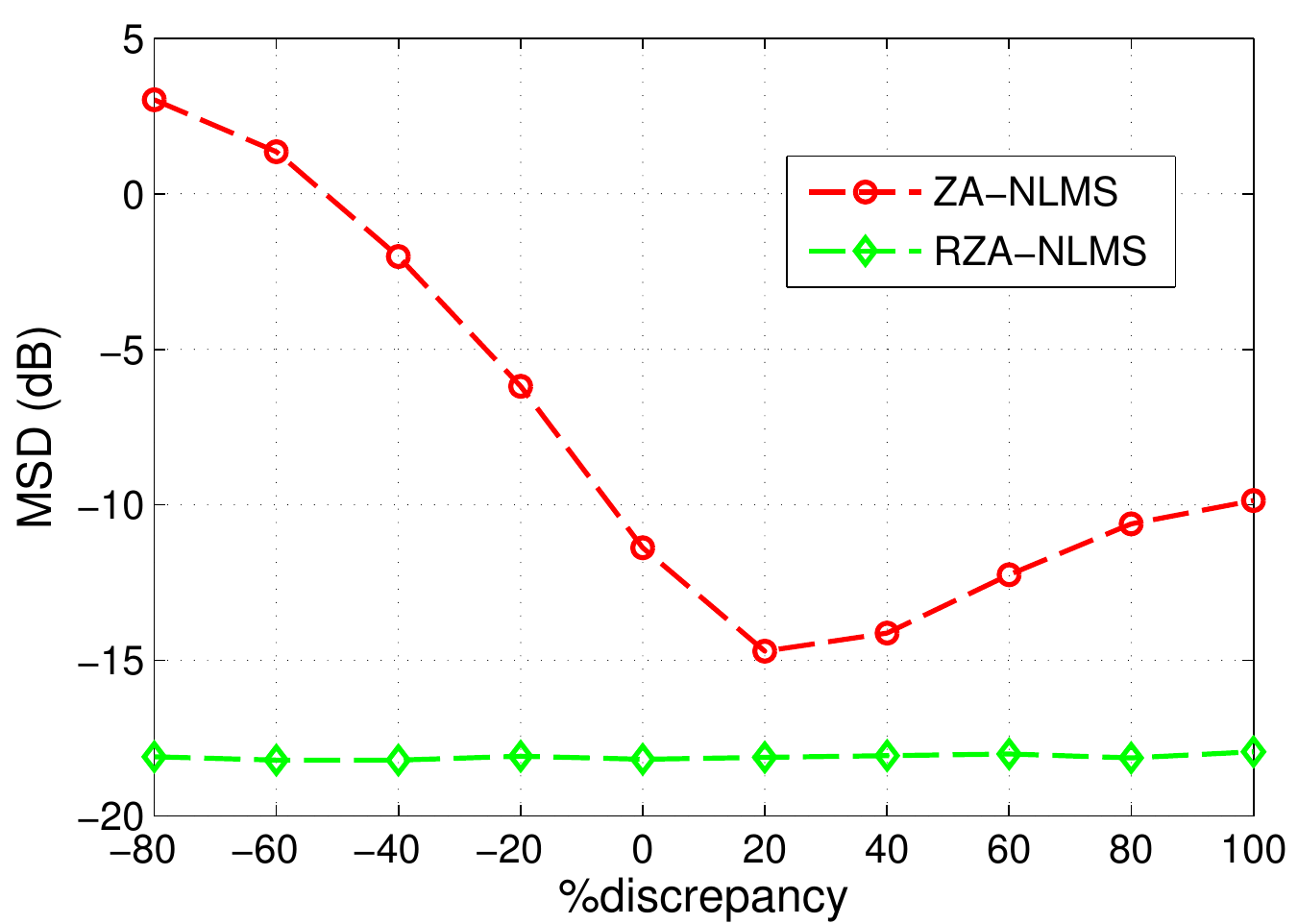}\\
  \caption{Sensitivity of ZA-NLMS and RZA-NLMS to $\eta_n$: MSD for ZA-NLMS and RZA-NLMS at the 750th iteration for white input signals.}\label{fig:5}
\end{center}
\end{figure}

Next, we investigate the sensitivity to $\eta_n$ for ZA-NLMS and RZA-NLMS. The result shown in Fig. \ref{fig:5} indicates that ZA-NLMS is more sensitive to $\eta_n$ than RZA-NLMS, which is highly robust to misspecified $\eta_n$.

\begin{figure}
\centering
\subfigure[]{
\includegraphics[scale=.5]{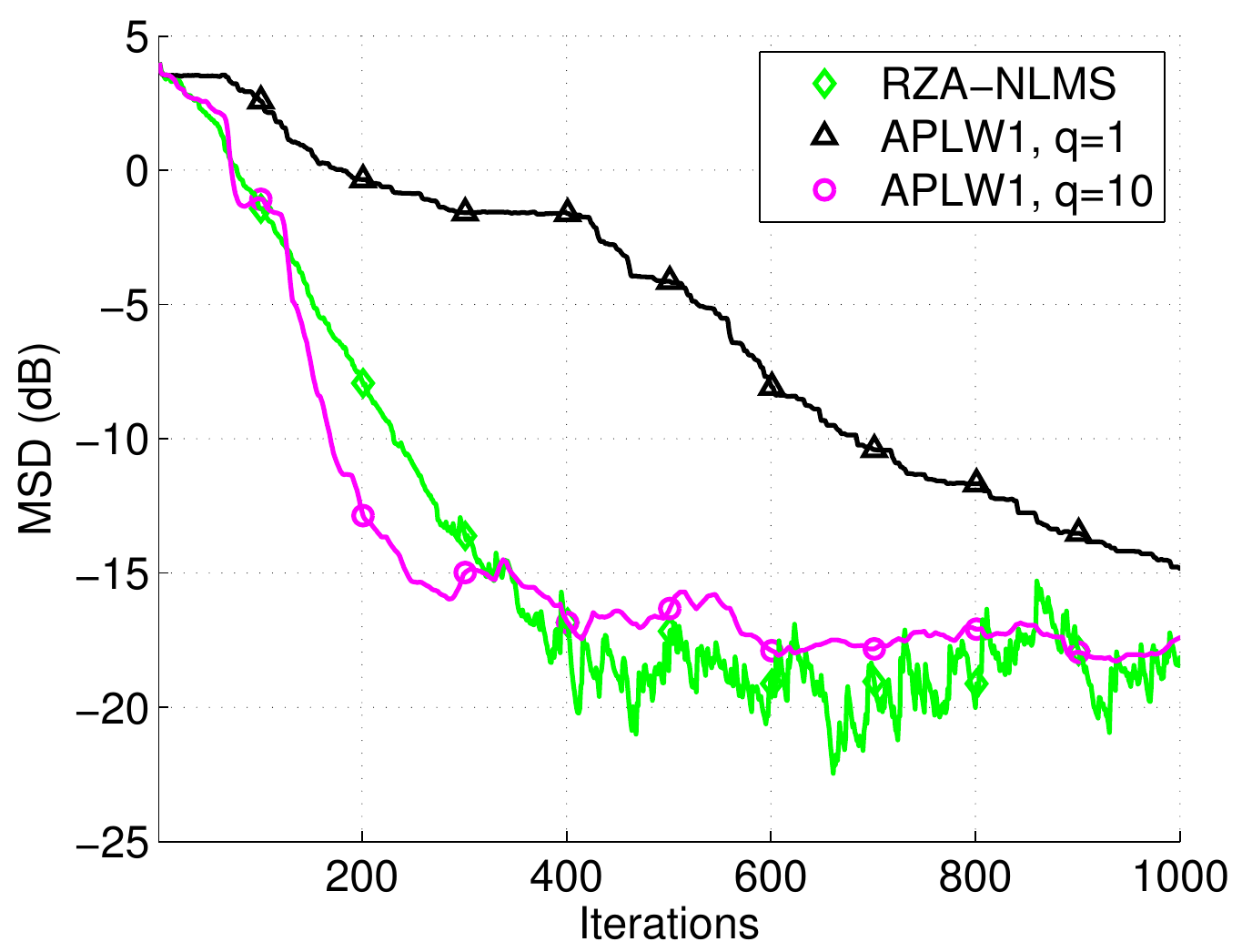}
}
\subfigure[]{
\includegraphics[scale=.5]{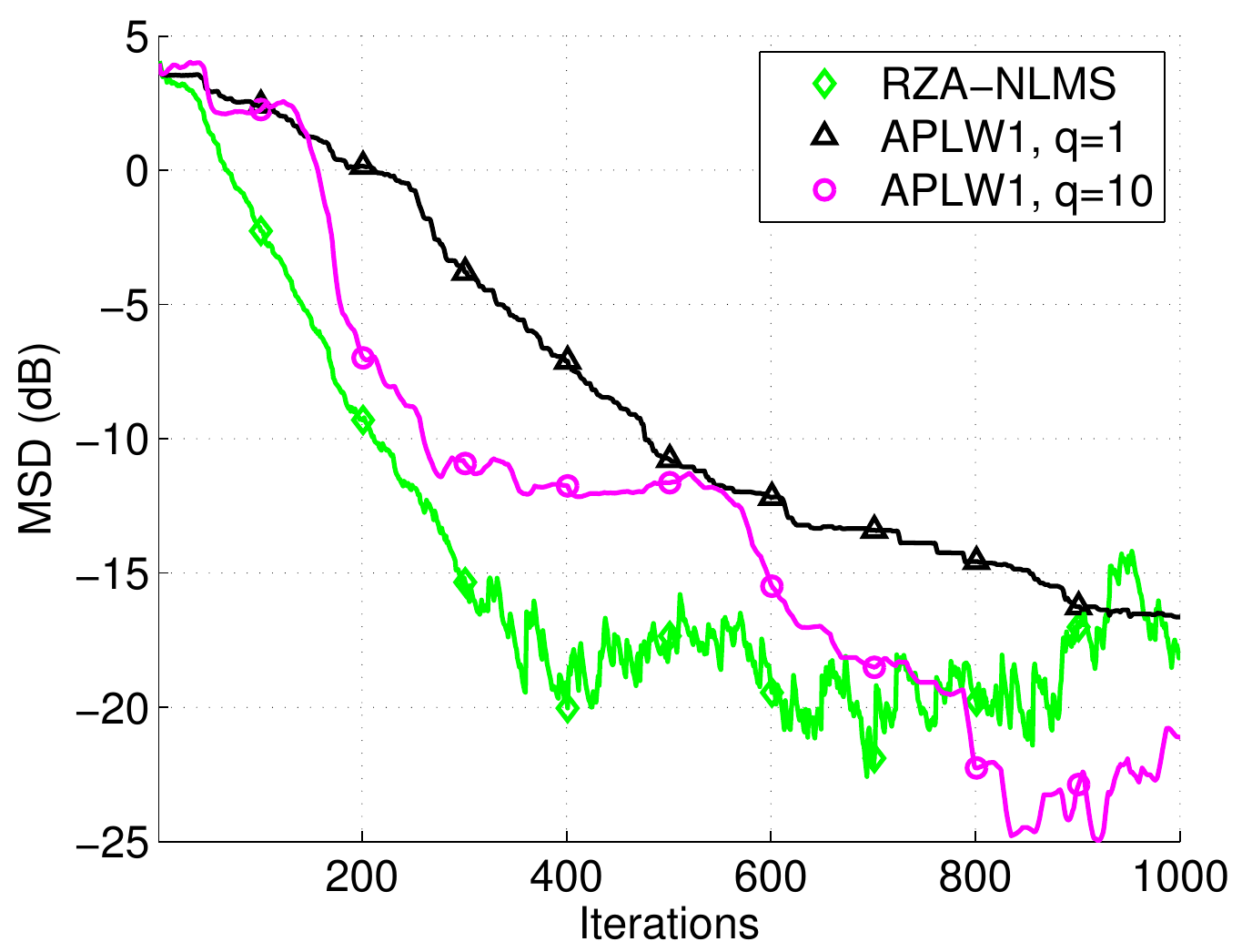}
}
\caption{Two different trials of RZA-NLMS and APWL1 for white input signals. APWL1 exhibits unstable convergence.}
\label{fig:6}
\end{figure}

Further analysis reveals that the projection based methods such APWL1 may exhibit unstable converging behaviors. Fig. \ref{fig:6} shows two independent trials of the simulation implemented in Fig. \ref{fig:4}. It can be seen that there exist several local minima in APWL1. For example, Fig. \ref{fig:6} (b) seems to indicate that APWL1 ($q = 10$) converges at the 400th iteration with MSD $\simeq -12$ dB, yet its MSD actually reaches values as low as $-25$ dB at the 900th iteration. This slow convergence  phenomenon is due to the fact that the weighted $\ell_1$ ball is determined in an online fashion and the projection operator is sensitive to mis-specifications of the convex set. In the contrast, our regularized LMS uses sub-gradient rather than projection to pursue sparsity, translating into improved convergence.

\subsubsection{Correlated input signals}
Next, we evaluate the filtering performance using correlated input signals. We generate the sequence $\{x_n\}$ as an AR(1) process
\begin{equation}
    \label{eq:colorinput}
  x_n = 0.8 x_{n-1} + u_n,
\end{equation}
which is then normalized to unit variance, where $\{u_n\}$ is a Gaussian i.i.d. process. The measurement system is the same as before and the variance of the noise is also $\sigma_v^2 = 0.1$.

\begin{figure}
\begin{center}
  \includegraphics[width=7cm]{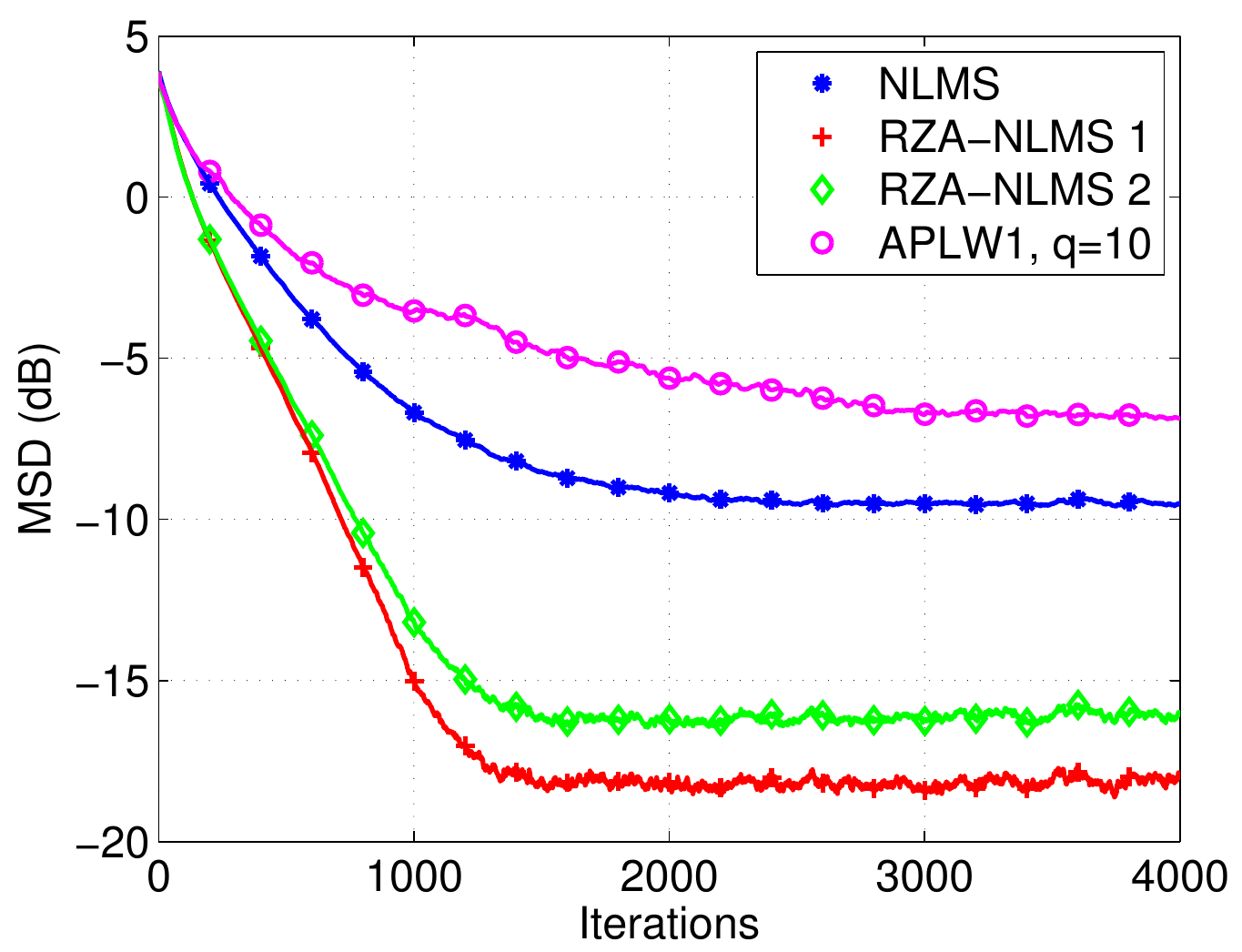}\\
  \caption{Correlated input signals: performance comparison for different filters, where RZA-NLMS 1 and RZA-NLMS 2 use (\ref{eq:rhonlms}) and (\ref{eq:wl12rho}) to determine $\rho_n$, respectively.}\label{fig:7}
\end{center}
\end{figure}

\begin{figure}
\centering
\subfigure[]{
\includegraphics[scale=.5]{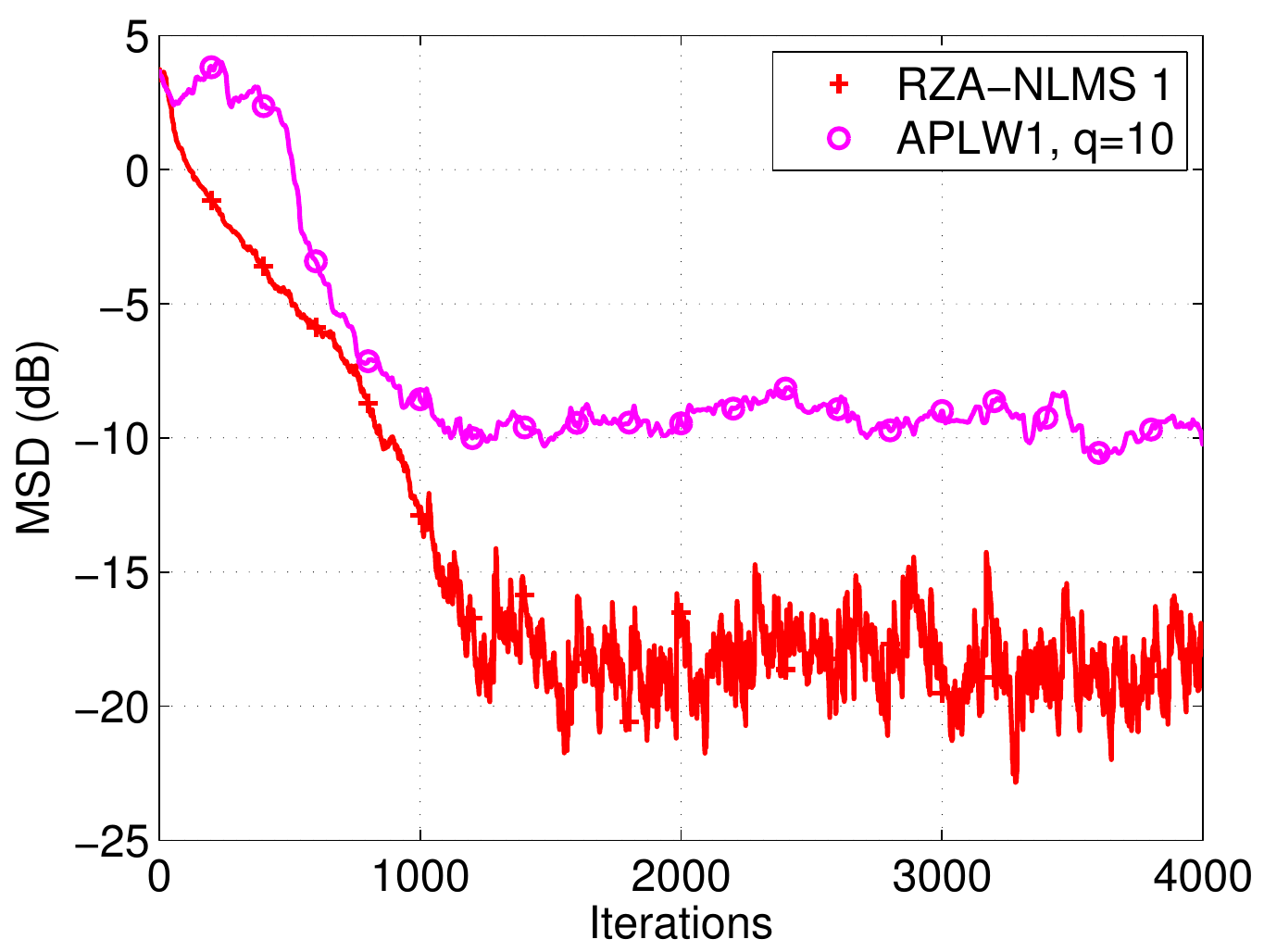}
}
\subfigure[]{
\includegraphics[scale=.5]{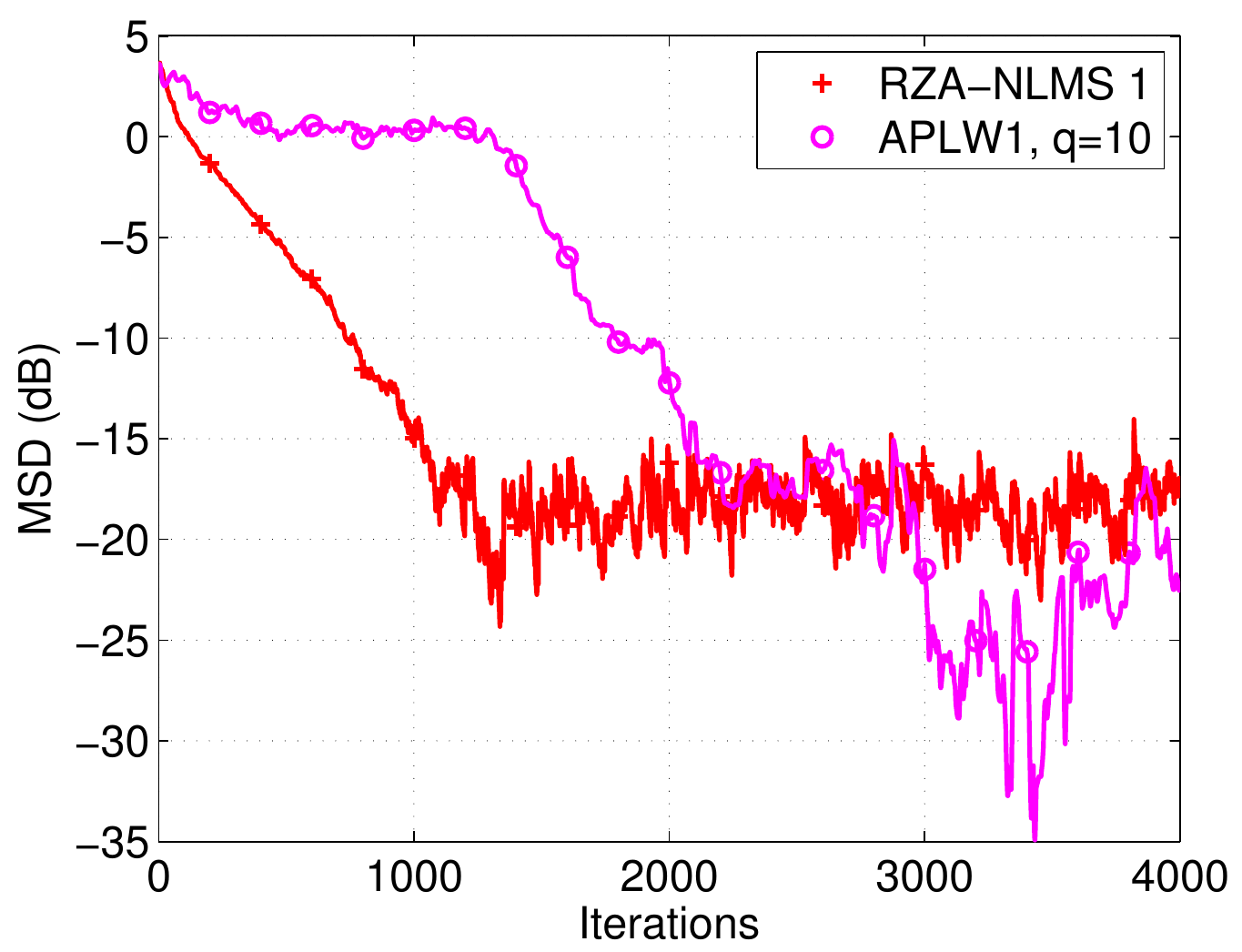}
}
\caption{Two different trials of RZA-NLMS and APWL1 for correlated input signals.}
\label{fig:8}
\end{figure}

We compare our RZA-NLMS with APWL1 ($q = 10$) and standard NLMS is also included as a benchmark. All the filter parameters are set to the same values as that in the previous simulation, except we employ both (\ref{eq:rhonlms}) and (\ref{eq:wl12rho}) to calculate $\rho_n$ in RZA-NLMS. The simulations are run 100 times and the average MSD curves are plotted in Fig. \ref{fig:7}. While Theorem \ref{thm:1} is derived based on white input assumptions, using (\ref{eq:rhonlms}) to determine $\rho_n$ achieves an empirically better performance compared to using (\ref{eq:wl12rho}) -- whose use guarantees dominance but yields a conservative result. This confirms our conjecture in Remark 7. We also observe a severe performance degradation of APWL1 for correlated input signals. Fig. \ref{fig:8} draws two independent trials in this simulation. The phenomenon described in Fig. \ref{fig:6} becomes more frequent when the input signal is correlated, which drags down the average performance of APWL1 significantly. Finally, we note that the filtering performance of a group sparse system (\eg Fig. \ref{fig:1} (b)) may be very different from that of a general sparse system. This will investigated in Section \ref{sec:groupsparse}.

\subsubsection{Tracking performance}
\begin{figure}
\begin{center}
  \includegraphics[width=7cm]{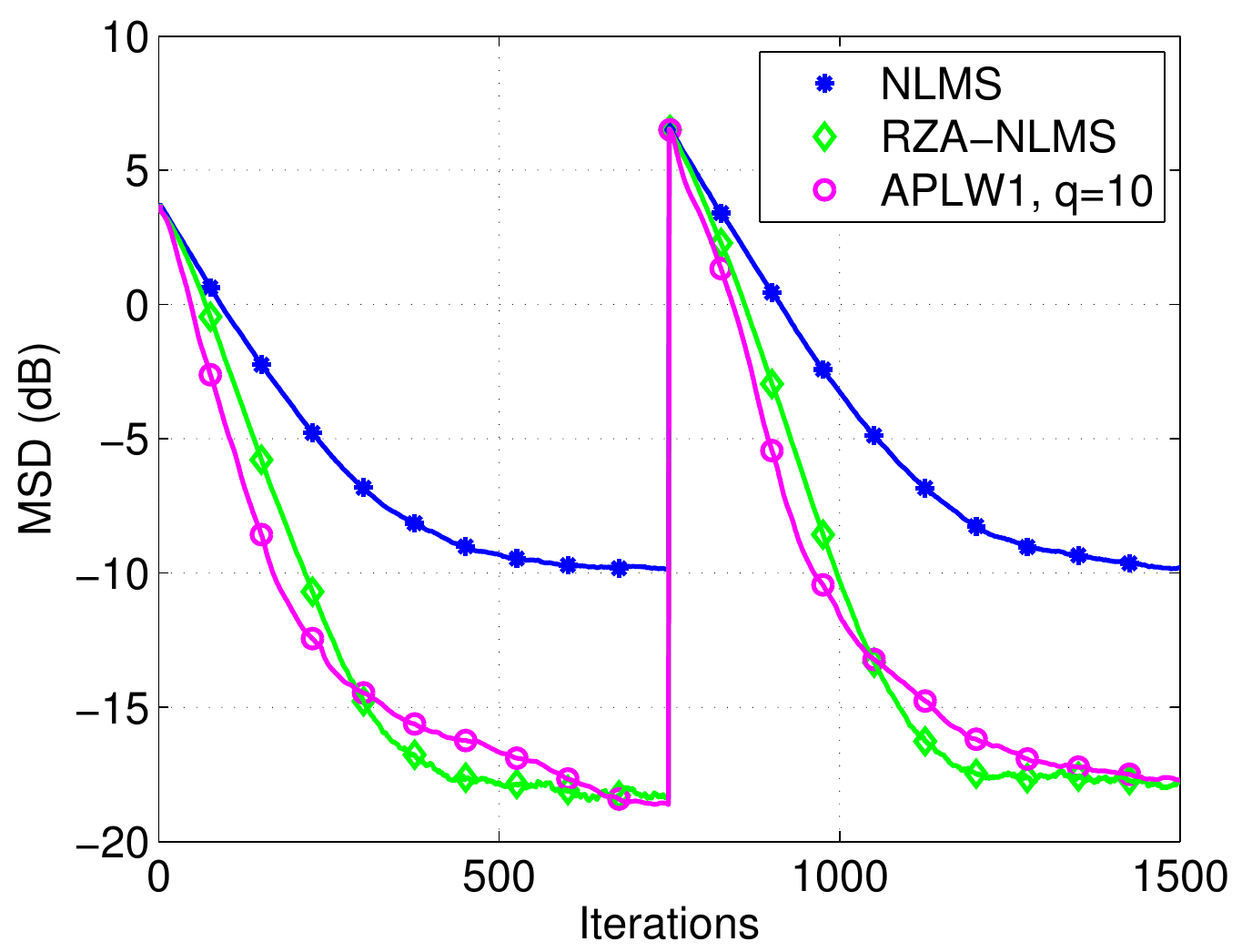}\\
  \caption{Comparison of tracking performances when the input signal is white.}\label{fig:9}
\end{center}
\end{figure}

Finally, we study the tracking performance of the proposed filters. The time-varying system is initialized using the same parameters as used to generate Fig. \ref{fig:3}. At the 750th iteration the system encounters a sudden change, where all the active coefficients are left-shifted for 10 taps. We use white input signals to excite the unknown system and all the filter parameters are set in an identical manner to Section \ref{sec:whiteinput}. The simulation is repeated  100 times and the averaged result is shown in Fig. \ref{fig:9}. It can be observed that both RZA-NLMS and APWL1 ($q=10$) achieve better tracking performance than the conventional NLMS.

\subsection{Identifying a group-sparse system}
\label{sec:groupsparse}
\begin{figure}
\begin{center}
  \includegraphics[width=7cm]{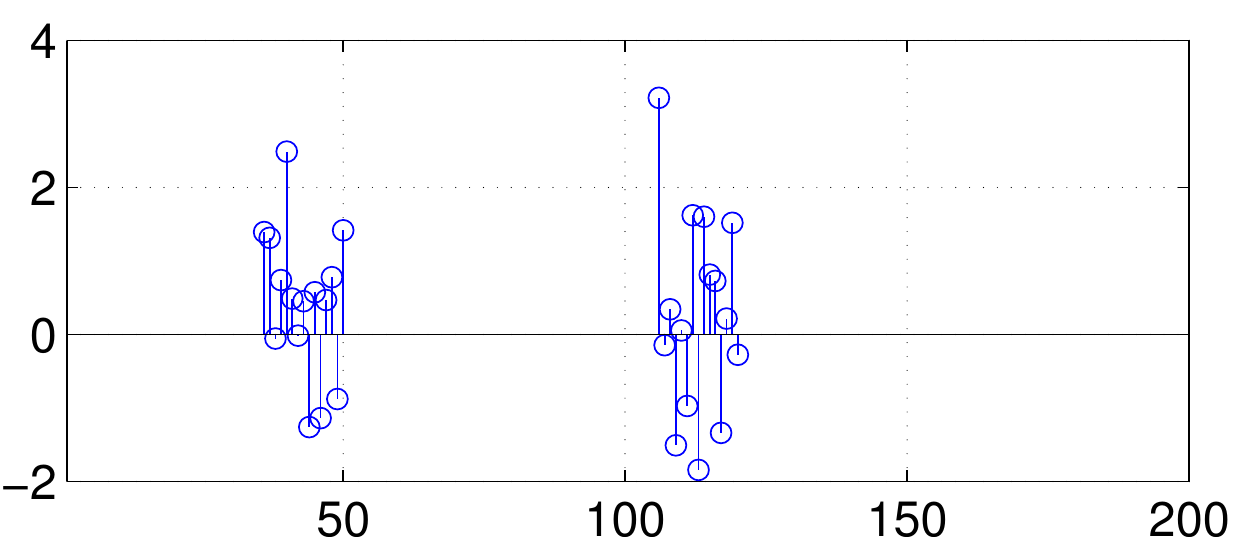}\\
  \caption{The group-sparse system used for simulations. There are two active blocks; each of them contains 15 non-zero coefficients.}\label{fig:10}
\end{center}
\end{figure}

Here we test performance of the group-sparse LMS filters developed in Section \ref{sec:groupsparselms}. The unknown system contains 200 coefficients that are distributed into two groups. The locations of the two groups are randomly selected, which start from the 36th tap and the 107th tap, respectively. Both of the two groups contain 15 coefficients and their values are randomly drawn from a standard Gaussian distribution. Fig. \ref{fig:10}  shows the response of the true system.

\begin{figure}
\begin{center}
  \includegraphics[width=7cm]{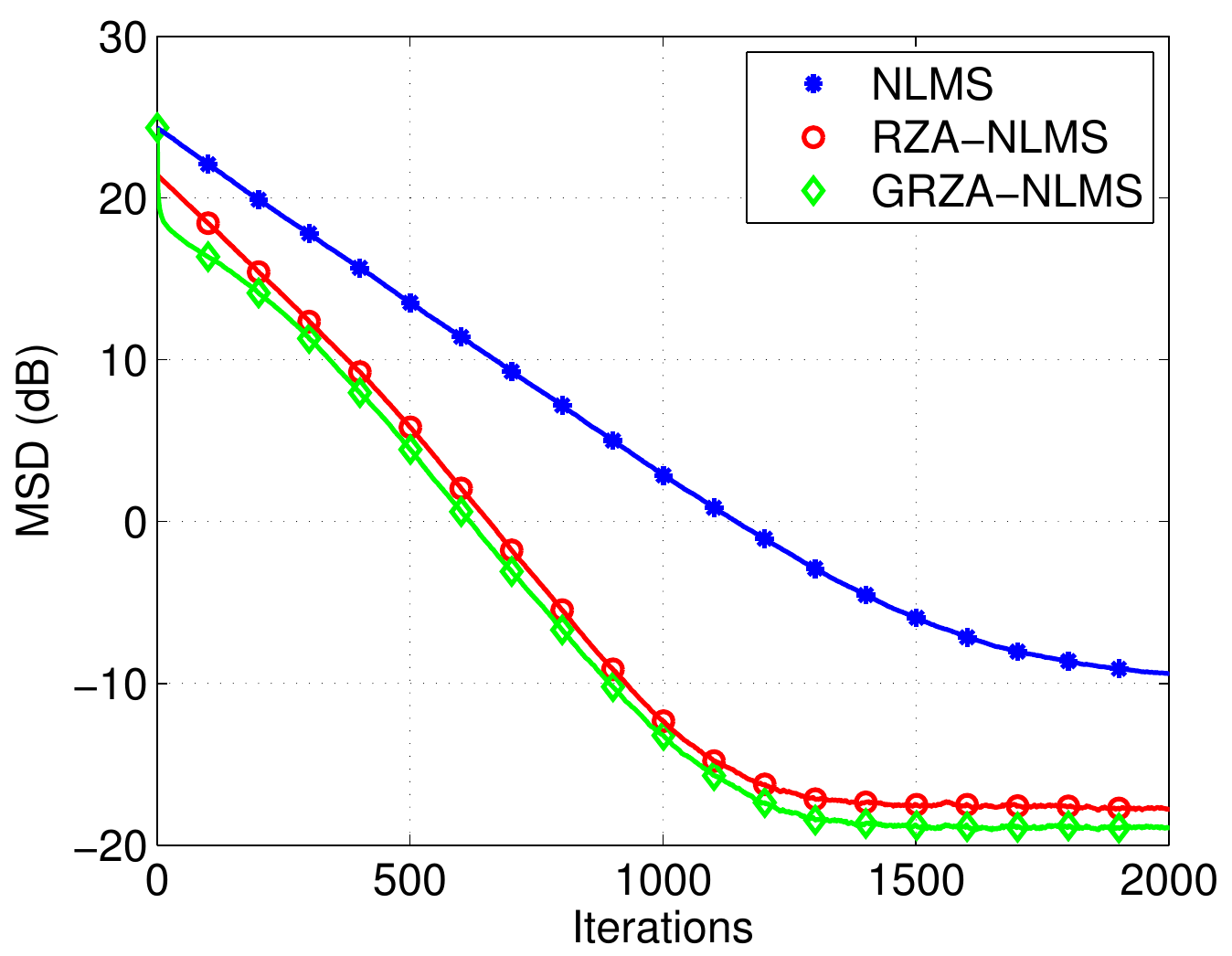}\\
  \caption{MSD comparison for the group-sparse system for white input signals.}\label{fig:11}
\end{center}
\end{figure}

The input signal $\{x_n\}$ is initially set to an i.i.d. Gaussian process and the variance of observation noise is $\sigma_v^2 = 0.1$. Three filters, GRZA-NLMS, RZA-NLMS and NLMS, are implemented, where the performance of NLMS is treated as a benchmark. In GRZA-NLMS, we divide the 200 coefficients equally into 20 groups, where each of them contains 10 coefficients. The step size $\mu_n$ of the three filters are all set according to (\ref{eq:nlms_stepsize}) with $\alpha_n = 1$. We use (\ref{eq:rhonlms}) to calculate $\rho_n$, where $\eta_n$ is set to 30 (the number of non-zero coefficients) for RZA-NLMS and 2 (the number of non-zero blocks) for GRZA-NLMS, respectively. We repeat the simulation 200 times and the averaged MSD is shown in Fig. \ref{fig:11}. It can be seen that GRZA-NLMS and RZA-NLMS outperform the standard NLMS for 10 dB in the steady-state MSD, while GRZA-NLMS only improves upon RZA-NLMS, but only marginally. This is partially due to the fact that in the white input scenario each coefficient is updated in an independent manner.

\begin{figure}
\begin{center}
  \includegraphics[width=7cm]{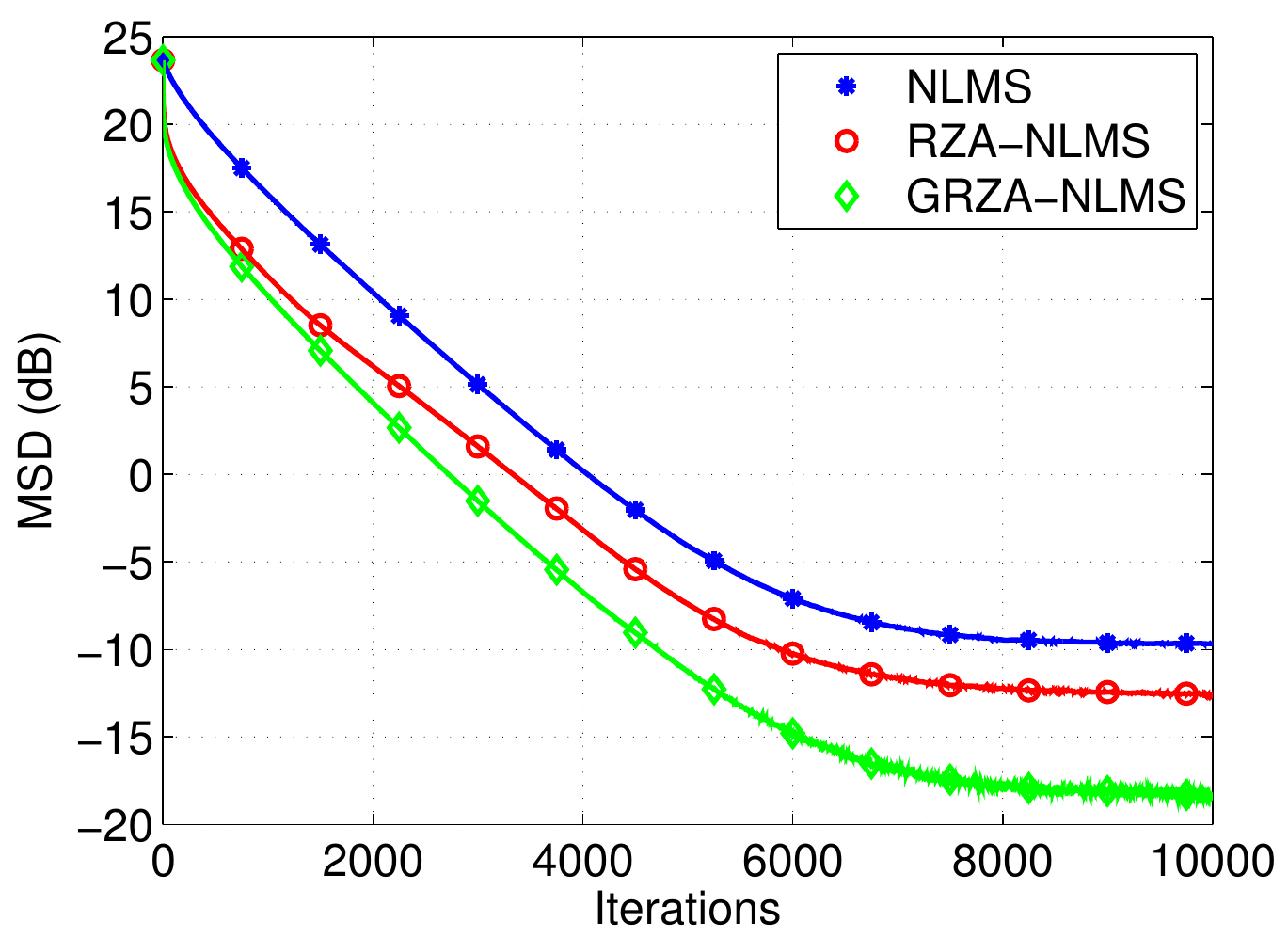}\\
  \caption{MSD comparison for the group-sparse system for correlated input signals.}\label{fig:12}
\end{center}
\end{figure}

We next consider the case of correlated input signals, where $\{x_n\}$ is generated by (\ref{eq:colorinput}) and then normalized to have unit variance. The parameters for all the filters are set to the same values as in the white input example and the averaged MSD curves are plotted in Fig. \ref{fig:12}. In the contrast to the white input example, here RZA-NLMS slightly outperforms NLMS but there is a significant improvement of GRZA-NLMS over RZA-NLMS. This demonstrates the power of promoting group-sparsity especially when the input signal is correlated.

\begin{figure}
\begin{center}
  \includegraphics[width=7cm]{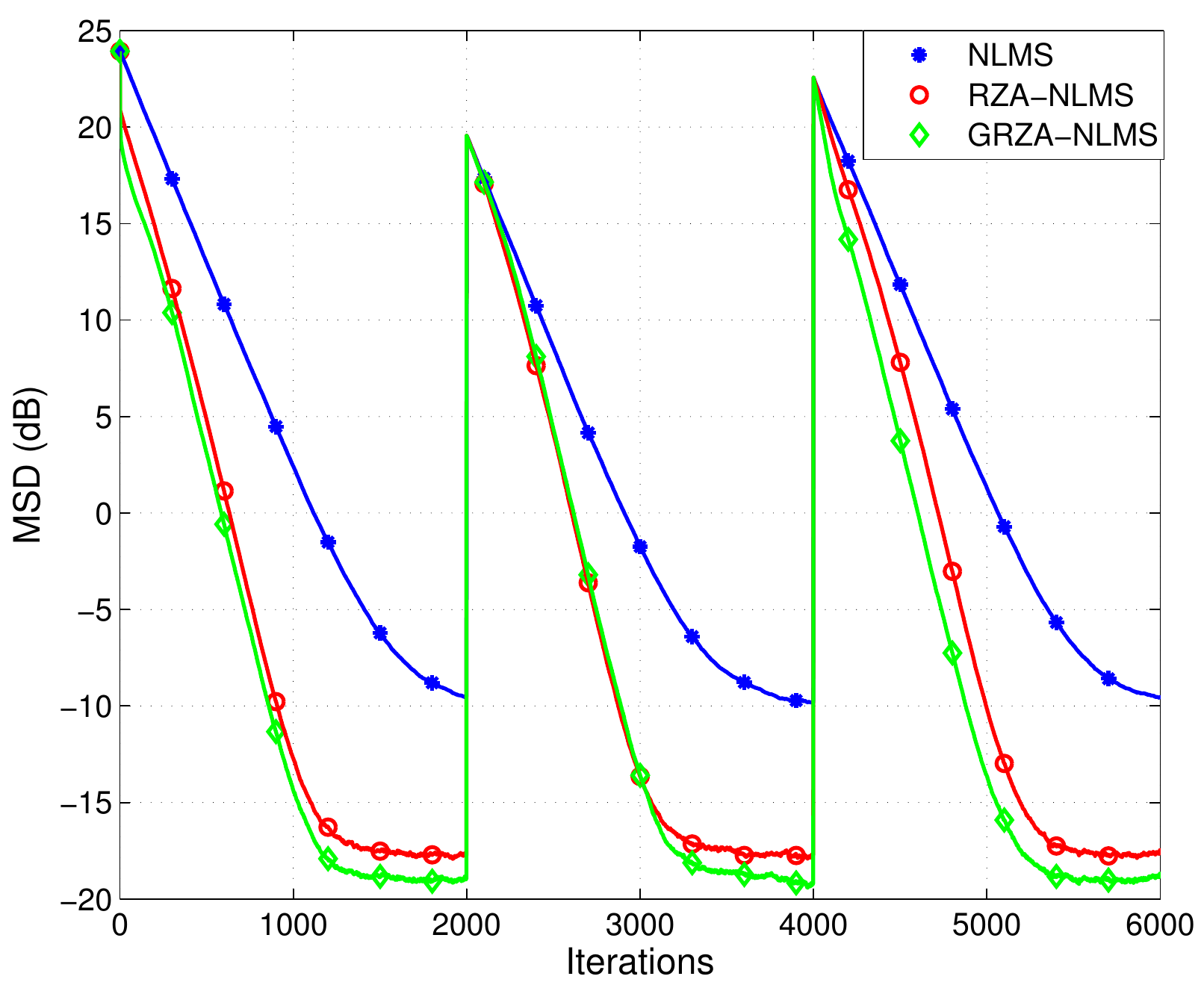}\\
  \caption{Tracking performance comparison for the group-sparse system for white input signals.}\label{fig:13}
\end{center}
\end{figure}

Finally, we evaluate the tracking performance of the adaptive filters. We use white signals as the system input and initialize the time-varying system using that in Fig. \ref{fig:10}. At the 2000th iteration, the system response is right-shifted for 50 taps, while the values of coefficients inside each block are unaltered. We then keep the block locations and reset the values of non-zero coefficients randomly at the 4000th iteration. From Fig. \ref{fig:13} we observe that the tracking rate of RZA-NLMS and GRZA-NLSM are comparable to each other when the system changes across blocks, and GRZA-NLMS shows a better tracking performance than RZA-NLMS when the system response changes only inside its active groups.

\section{Conclusion}
In this paper we proposed a general class of LMS-type filters regularized by convex sparsifying penalties. We derived closed-form expressions for choosing the regularization parameter that guarantees provable dominance over conventional LMS filters. We applied the proposed regularized LMS filters to sparse and group-sparse system identification and demonstrated their performances using numerical simulations.

Our regularized LMS filter is derived from the LMS framework and inherits its simplicity, low computational cost and low memory requirements, and robustness to parameter mismatch. It is likely that the convergence rate and steady-state performance can be improved by extension to second-order methods, such as RLS and Kalman filters. Efficient extensions of our results for sparse/group-sparse RLS filters are a worthy topic of future study.

\section{Appendix}
\subsection{Proof of Theorem \ref{thm:1}}
We prove Theorem \ref{thm:1} for LMS, \ie the case that $\mu_n$ are constants. NLMS, where $\mu_n$ is determined by (\ref{eq:nlms_stepsize}), can be derived in a similar manner.

According to (\ref{eq:rlms}),
\begin{equation}
    \label{eq:th1_32}
\begin{aligned}
& \hw_{n+1}-\bw \\
& = (\bbi-\mu_n\bx_n\bx_n^T)(\hw_n-\bw)-\rho_n
\partial f_n(\hw_n)+\mu_n v_n \bx_n.
\end{aligned}
\end{equation}
Noting that $\hw_n$, $\bx_n$ and $v_n$ are mutually independent,
we have
\small
\begin{equation}
\label{eq:smsd_sublms1}
\begin{aligned}
    & E\blc\|\hw_{n+1}-\bw\|^2|\hw_n\brc = \\
    & (\hw_n-\bw)^T
    E\blc\bl\bbi-\mu_n\bx_n\bx_n^T\br^2\brc(\hw_n-\bw) + \mu_n^2 \sigma_v^2E\blc\|\bx_n\|^2\brc\\
    & +2\rho_n (\bw-\hw_n)^T E\blc\bbi-\mu_n\bx_n\bx_n^T\brc\partial
    f_n(\hw_n) +\rho_n^2 \|\partial f_n(\hw_n)\|^2.
\end{aligned}
\end{equation}
\normalsize
As $\{x_n\}$ is a Gaussian i.i.d. process, $\bx_n$ is a Gaussian random vector with mean zero and covariance $\sigma_x^2\bbi$. Thus,
\begin{equation}
    \label{eq:th1_33}
  E\blc\bl\bbi-\mu_n\bx_n\bx_n^T\br^2\brc = (1-2\sigma_x^2\mu_n+N\sigma_x^4\mu_n^2)\bbi,
\end{equation}
\begin{equation}
    \label{eq:th1_34}
  E\blc\bbi-\mu_n\bx_n\bx_n^T\brc = (1-\sigma_x^2\mu_n)\bbi,
\end{equation}
and
\begin{equation}
    \label{eq:th1_35}
  E\blc\|\bx_n\|^2\brc = N\sigma_x^2.
\end{equation}
Substituting (\ref{eq:th1_33}), (\ref{eq:th1_34}) and (\ref{eq:th1_35}) into (\ref{eq:smsd_sublms1}), we have
\begin{equation}
\label{eq:smsd_sublms2}
\begin{aligned}
    & E\blc\|\hw_{n+1}-\bw\|^2|\hw_n\brc = \\
    & (1-2\sigma_x^2\mu_n+N\sigma_x^4\mu_n^2) ~ \|\hw_{n}-\bw\|^2 + N \mu_n^2 \sigma_x^2 \sigma_v^2\\
    & +2\rho_n (1-\sigma_x^2\mu_n)(\bw-\hw_n)^T \partial
    f_n(\hw_n) +\rho_n^2 \|\partial f_n(\hw_n)\|^2.
\end{aligned}
\end{equation}
As $f_n(\cdot)$ is a convex function, by the definition of sub-gradient, we have
\begin{equation}
    \label{eq:th1_38}
(\bw-\hw_n)^T \partial f_n(\hw_n) \le f_n(\bw) - f_n(\hw_n) \le \eta_n - f_n(\hw_n).
\end{equation}
Therefore,
\begin{equation}
\label{eq:smsd_sublms3}
\begin{aligned}
    & E\blc\|\hw_{n+1}-\bw\|^2|\hw_n\brc \le \\
    & (1-2\sigma_x^2\mu_n+N\sigma_x^4\mu_n^2) ~ \|\hw_{n}-\bw\|^2 + N \mu_n^2 \sigma_x^2 \sigma_v^2\\
    & - 2\rho_n (1-\sigma_x^2\mu_n)(f_n(\hw_n)-\eta_n) +\rho_n^2 \|\partial f_n(\hw_n)\|^2.
\end{aligned}
\end{equation}
Define
\begin{equation}
  C(\rho_n) = - 2\rho_n (1-\sigma_x^2\mu_n)(f_n(\hw_n)-\eta_n) +\rho_n^2 \|\partial f_n(\hw_n)\|^2,
\end{equation}
and take expectation on both sides of (\ref{eq:smsd_sublms3}) with respect to $\hw_n$ to obtain
\begin{equation}
    \begin{aligned}
  & E\blc \|\hw_{n+1}-\bw\|^2\brc \\
  & \le (1-2\sigma_x^2\mu_n+N\sigma_x^4\mu_n^2) E\blc\|\hw_{n}-\bw\|^2\brc + N \mu_n^2 \sigma_x^2 \sigma_v^2 \\
  & \quad + E\blc C(\rho_n)\brc.
    \end{aligned}
\end{equation}
It is easy to check that $C(\rho_n) \le 0$ if $\rho_n \in [0, 2\rho_n^*]$, where $\rho_n^*$ is defined in (\ref{eq:rholms}). Therefore,
\begin{equation}
    \label{eq:th1_42}
    \begin{aligned}
  & E\blc \|\hw_{n+1}-\bw\|^2\brc \\
  & \le (1-2\sigma_x^2\mu_n+N\sigma_x^4\mu_n^2) E\blc\|\hw_{n}-\bw\|^2\brc + N \mu_n^2 \sigma_x^2 \sigma_v^2 \\
    \end{aligned}
\end{equation}
if $\rho_n \in [0,2\rho_n^*]$.
For the standard LMS, there is
\begin{equation}
    \label{eq:th1_43}
    \begin{aligned}
  & E\blc \|\hw'_{n+1}-\bw\|^2\brc \\
  & = (1-2\sigma_x^2\mu_n+N\sigma_x^4\mu_n^2) E\blc\|\hw_{n}'-\bw\|^2\brc + N \mu_n^2 \sigma_x^2 \sigma_v^2.
    \end{aligned}
\end{equation}
Therefore, under the condition that $ E\blc \|\hw_0-\bw\|^2\brc = E\blc \|\hw_0'-\bw\|^2\brc $, (\ref{eq:strongdominance}) can be obtained from (\ref{eq:th1_42}) and (\ref{eq:th1_43}) using a simple induction argument.

\subsection{Proof of Theorem \ref{thm:2}}
We start our proof from (\ref{eq:th1_32}) and calculate the following conditional MSD:
\begin{equation}
  \begin{aligned}
  & E\blc\|\hw_{n+1}-\bw\|^2|\hw_n,\bx_n\brc = \\
  & (\hw_n-\bw)^T(\bbi-\mu_n \bx_n\bx_n^T)^2(\hw_n-\bw)+ \mu_n^2\sigma_v^2 \|\bx_n\|^2+D(\rho_n),
  \end{aligned}
\end{equation}
where
\begin{equation}
    \label{eq:th2_45}
 D(\rho_n) =  2\rho_n (\bw-\hw_n)^T(\bbi-\mu_n \bx_n\bx_n^T)\partial f_n(\hw_n)+\rho_n^2\|\partial f_n(\hw_n)\|^2.
\end{equation}
For the cross term $2\rho_n (\bw-\hw_n)^T(\bbi-\mu_n \bx_n\bx_n^T)\partial f_n(\hw_n)$ we have
\begin{equation}
    \label{eq:th2_46}
\begin{aligned}
  & 2\rho_n(\bw-\hw_n)^T(\bbi-\mu_n \bx_n\bx_n^T)\partial f_n(\hw_n)\\
  & = 2\rho_n(\bw-\hw_n)^T\partial f_n(\hw_n) + 2\rho_n\mu_n\hw_n^T\bx_n \cdot \bx_n^T\partial f_n(\hw_n)\\
  & \quad - 2\rho_n\mu_n\bw^T\bx_n \cdot \bx_n^T\partial f_n(\hw_n)\\
  & \le 2\rho_n(\eta_n - f_n(\hw_n))+ 2\rho_n\mu_n\hw_n^T\bx_n \cdot \bx_n^T\partial f_n(\hw_n) \\
  & \quad + 2\rho_n \mu_n\left|\bw^T\bx_n\right|\cdot \left|\bx_n^T\partial f_n(\hw_n)\right|.
\end{aligned}
\end{equation}
We now establish upper-bounds for $|\bw^T\bx_n|$. Indeed,
\begin{equation}
    \label{eq:th2_47}
\begin{aligned}
\left|\bw^T\bx_n\right| & = \left|\sum_{j=1}^J \bw_{I_j}^T\bx_{n,I_j}\right|\\
&\le \sum_{j=1}^J \left|\beta_{n,j}\bw_{I_j}^T\frac{1}{\beta_{n,j}}\bx_{n,I_j}\right|\\
& \le \sum_{j=1}^J \beta_{n,j}\|\bw_{I_j}\|_2 \frac{\|\bx_{n,I_j}\|_2}{\beta_{n,j}}\\
& \le \blc\sum_{j=1}^J \beta_{n,j}\|\bw_{I_j}\|_2\brc \max_j \frac{\|\bx_{n,I_j}\|_2}{\beta_{n,j}}\\
& = f_n(\bw_n)\max_j \frac{\|\bx_{n,I_j}\|_2}{\beta_{n,j}} \le \eta_n \max_j \frac{\|\bx_{n,I_j}\|_2}{\beta_{n,j}}.
\end{aligned}
\end{equation}
Substituting (\ref{eq:th2_46}) and (\ref{eq:th2_47}) into (\ref{eq:th2_45}) we obtain that
\begin{equation}
D(\rho_n) \le -2\rho_n(f_n(\hw_n)-\eta_n-\mu_n r_n) + \rho_n^2\|\partial f_n(\hw_n)\|_2^2,
\end{equation}
where $r_n$ is defined in (\ref{eq:def_rn}). Note that $D(\rho_n) \le 0$ if $\rho_n \in [0,2\rho_n^*]$ ($\rho_n^*$ is defined in (\ref{eq:wl12rho})). There is
\begin{equation}
  \begin{aligned}
  & E\blc\|\hw_{n+1}-\bw\|^2|\hw_n,\bx_n\brc  \\
  & \le (\hw_n-\bw)^T(\bbi-\mu_n \bx_n\bx_n^T)^2(\hw_n-\bw)+ \mu_n^2\sigma_v^2 \|\bx_n\|^2,
  \end{aligned}
\end{equation}
if $\rho_n \in [0,2\rho_n^*]$. Therefore,
\begin{equation}
  \begin{aligned}
  & E\blc\|\hw_{n+1}-\bw\|^2|\brc  \\
  & \le E\blc(\hw_n-\bw)^T(\bbi-\mu_n \bx_n\bx_n^T)^2(\hw_n-\bw)\brc\\
  & \quad + \mu_n^2\sigma_v^2 E\blc\|\bx_n\|^2\brc\\
  & = E\blc(\hw_n'-\bw)^T(\bbi-\mu_n \bx_n\bx_n^T)^2(\hw_n'-\bw)\brc\\
  & \quad + \mu_n^2\sigma_v^2 E\blc\|\bx_n\|^2\brc\\
  & = E\blc\|\hw_{n+1}'-\bw\|^2|\brc,
  \end{aligned}
\end{equation}
which proves Theorem \ref{thm:2}.

\bibliographystyle{IEEEbib}
\bibliography{strings,refs}

\end{document}